\documentclass{ecai} 

\usepackage{latexsym}
\usepackage{amssymb}
\usepackage{amsmath}
\usepackage{amsthm}
\usepackage{booktabs}
\usepackage{enumitem}
\usepackage{graphicx}
\usepackage{color}
\usepackage{tabularx}
\usepackage{enumitem}
\usepackage{array}
\usepackage{algorithmic}
\usepackage[linesnumbered, ruled, vlined]{algorithm2e}

\newcommand{\BibTeX}{B\kern-.05em{\sc i\kern-.025em b}\kern-.08em\TeX}

\begin{document}


\begin{frontmatter}


\paperid{123} 


\title{Leveraging User-Generated Reviews for Recommender Systems with Dynamic Headers}

\author[A]{\fnms{Shanu}~\snm{Vashishtha}\thanks{Corresponding Author. Email: Shanu.Vashishtha@walmart.com}\footnote{Equal contribution.}}
\author[A]{\fnms{Abhay}~\snm{Kumar}\footnotemark}
\author[A]{\fnms{Lalitesh}~\snm{Morishetti}} 
\author[A]{\fnms{Kaushiki}~\snm{Nag}} 
\author[A]{\fnms{Kannan}~\snm{Achan}} 

\address[A]{Personalization and Recommendation,  Walmart Global Tech, Sunnyvale, USA}




\begin{abstract}
E-commerce platforms have a vast catalog of items to cater to their customers’ shopping interests. Most of these platforms assist their customers in the shopping process by offering optimized recommendation carousels, designed to help customers quickly locate their desired items. Many models have been proposed in academic literature to generate and enhance the ranking and recall set of items in these carousels. Conventionally, the accompanying carousel title text (header) of these carousels remains static. In most instances, a generic text such as "Items similar to your current viewing" is utilized. Fixed variations such as the inclusion of specific attributes "Other items from a similar seller" or "Items from a similar brand" in addition to "frequently bought together" or "considered together" are observed as well. This work proposes a novel approach to customize the header generation process of these carousels. Our work leverages user-generated reviews that lay focus on specific attributes (aspects) of an item that were favorably perceived by users during their interaction with the given item. We extract these aspects from reviews and train a graph neural network-based model under the framework of a conditional ranking task. We refer to our innovative methodology as Dynamic Text Snippets (DTS) which generates multiple header texts for an anchor item and its recall set. Our approach demonstrates the potential of utilizing user-generated reviews and presents a unique paradigm for exploring increasingly context-aware recommendation systems. 
\end{abstract}

\end{frontmatter}


\section{Introduction}
A recommender system, in the era of e-commerce, serves as a critical tool that offers suggestions personalized to users' tastes and preferences. These systems employ machine learning algorithms to analyze users' behavior and preferences, thereby providing accurate recommendations\cite{zhang2019deep}. The immense value of recommender systems in improving customer experience, enhancing user engagement, and boosting businesses' profitability is undeniable \citep{jannach2019measuring, tarnowska2020recommender}. However, a specific yet significant challenge that arises in the context of these systems is generating dynamic header text for the accompanying carousel in which recommendations are shown to the users. This task involves crafting unique and appealing headers that not only capture the essence of the recommended content but also resonate with individual user preferences. 

Generating dynamic header text for a recommendation carousel presents significant challenges that requires a nuanced understanding of both the content and user behaviour. Firstly, capturing user preferences towards an item is a complex task due to the individual variability and subjectivity involved in preferences, which are often influenced by cognitive, emotional and situational factors. Secondly, accurately providing immediate context and ensuring a direct correlation between the headers and carousel items is necessary for an improved navigation, user engagement, and overall coherence of the digital interface. Lastly, these dynamic headers must positively appeal to users as they serve as the initial point of interaction, influencing user engagement with the carousel items and can significantly impact user satisfaction and retention rates. If these headers are not appealing, user attention may drift leading to decreased interaction, even if the recommendation items are relevant. All these reasons necessitate the development of sophisticated algorithms and robust underlying data analysis techniques demanding a scalable solution. 

In this paper, we explore the under examined but promising impact of dynamic header texts. This unique approach can potentially optimize the functionality of recommender systems by offering more tailored and relevant user experiences. A crucial aspect to consider is the provision of straightforward explanations to users. When users understand why certain content is shown to them, their confidence increases, thereby potentially boosting their overall engagement and satisfaction. It offers a novel yet subtle way to potentially improve key business metrics. The benefits of addressing this problem, therefore, are multifaceted and provide a fresh perspective on user interface personalization. 

There is a scarcity of literature specifically focused on dynamic header text in recommender systems. However, we came across some related studies in the broader field of personalized user experience. For instance, \citet{mcnee2006being} discusses the importance of explaining recommendations to users. Another research \citep{konstan2012recommender} also discusses the potential benefits of personalization and adaptivity in recommender systems. But most of these were focused only on the items shown to the users while none of these studies explicitly address the creation of dynamic header text for recommendations, thus marking an unexplored avenue in the research landscape. Our work aims to bridge this gap by investigating the experiential aspect of dynamic header text. We believe that understanding how this feature can enhance users' confidence and satisfaction levels could provide valuable insights for achieving a personalized header text offering explainable recommendations to users. 

\begin{figure*}[ht]
  \centering
  \begin{tabular}{@{}c@{\hspace{1mm}}c@{}}
    \includegraphics[width=0.5\textwidth]{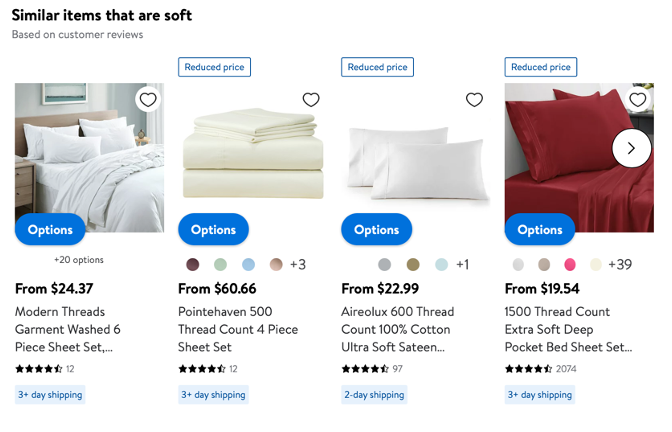} &
    \includegraphics[width=0.5\textwidth]{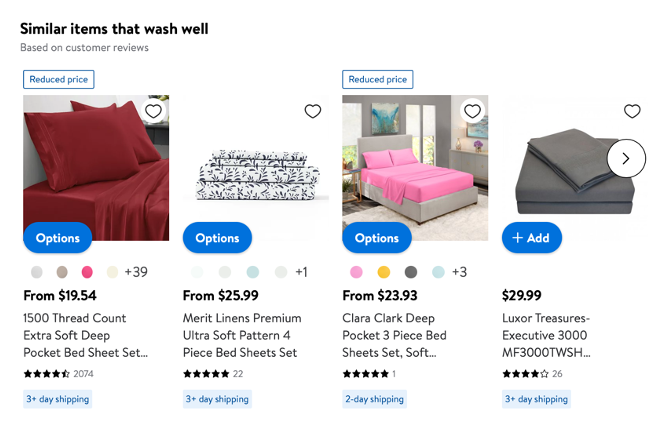} \\
    \includegraphics[width=0.5\textwidth]{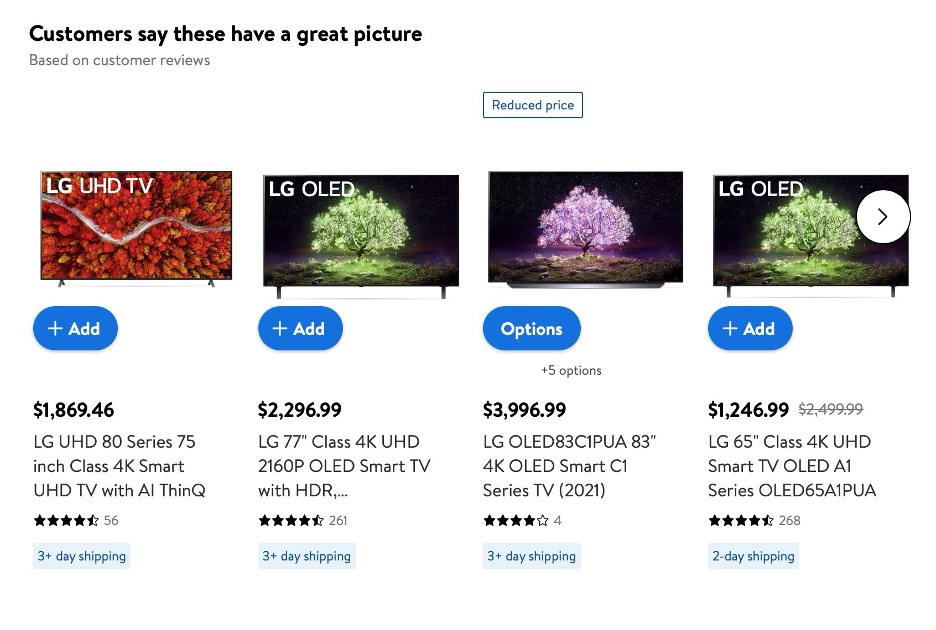} &
    \includegraphics[width=0.5\textwidth]{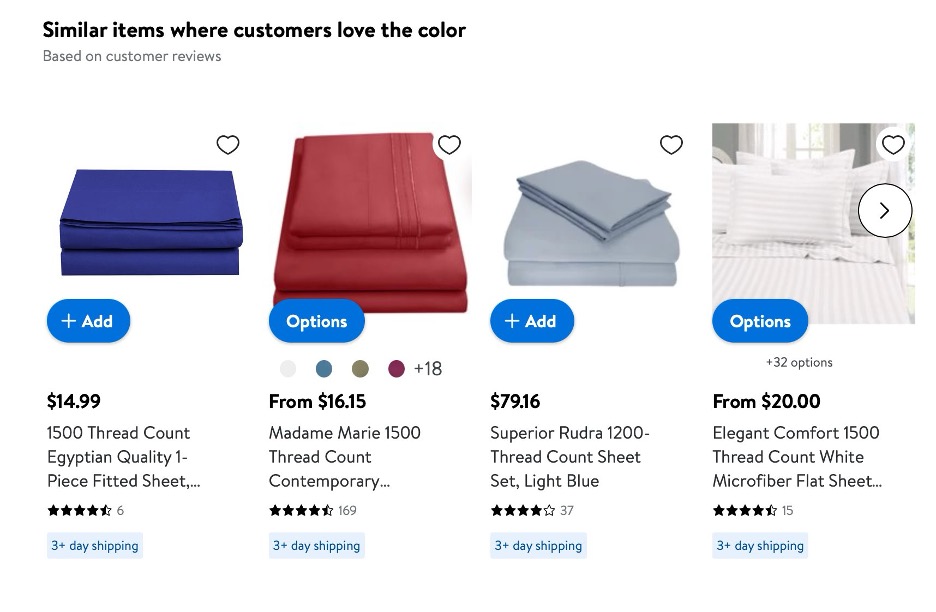} \\
  \end{tabular}
  \caption{Examples of Dynamic Text Snippets (DTS) carousel for e-commerce recommendations. Each carousel has recommended items with shared aspects and the associated explainable header.}
  \label{fig:your_label}
\end{figure*}

In this work, we propose a novel aspect-based recommender system - Dynamic Text Snippets (DTS) that harnesses the power of user-generated reviews to identify and link items through shared aspects. Our approach involves extracting aspect information for items from user reviews and incorporating it into the model training phase, allowing for a more contextualized and nuanced understanding of user preferences. The aspect extraction includes a preprocessing stage followed by candidate aspect identification. Next, we train a graph neural network-based recommendation model that leverages this aspect information to generate relevant items with multiple header texts for a given set of reference items (anchor items). To evaluate the effectiveness of our proposed DTS model, we conduct a series of offline and online experiments. Our results demonstrate that the incorporation of aspect information leads to significant improvements in the diversity of recommendations.   



\section{Background and Related Work}
\subsection{Graph Neural Networks }
E-commerce recommender systems have been an area of significant interest over the past few years, owing to their potential to drive customer engagement and sales \citep{jannach2019measuring}. Advances in machine learning and the consideration of contextual information have further improved the quality and personalization of recommendations. \citet{zhang2019deep} proposed a deep learning-based recommender system that combines user-item interactions and user/item metadata for recommendation, demonstrating promising results. Recently, graph neural network (GNN) techniques have been widely utilized in recommender systems \cite{10.1145/3535101}.
This trend stems from the inherent graph-structured nature of recommendation system data and the effectiveness of GNNs in learning informative representations from graphs.

The GNNs learn expressive representations from graph-structured data by effectively capturing high-order connectivity, going beyond direct user-item interactions to uncover complex patterns. \cite{10.1145/3535101} They also exploit the structural properties of the recommendation graph to learn richer representations and offer the flexibility to seamlessly integrate side information (such as item attributes or social relationships) to further refine recommendations. The inherent graph structure of recommender systems, where users and items form nodes with interactions as edges (capturing interactions like clicks, add-to-carts, and transactions), aligns naturally with the capabilities of GNNs. 

Graph Attention Networks (GATs) \citep{velivckovic2017graph} are a significant advancement in graph neural networks, introducing attention mechanisms that enable the model to dynamically learn the importance of different neighbors within a graph structure. This is achieved by assigning learnable attention coefficients to each edge, determining how much a neighbor's features should influence a node's updated representation.  GATs offer advantages like adaptive focusing on the most relevant parts of the graph, leading to improved representations. Additionally, they automate the process of determining neighbor importance, reducing the need for manual feature engineering.

Building on the foundation of GATs, \citet{cen2019representation} addresses the challenge of learning meaningful node embeddings within complex network structures which are common in real-world applications like e-commerce. Specifically, it focuses on Attributed Multiplex Heterogeneous Networks (AMHENs) where nodes possess diverse types, attributes, and participate in multiple layers based on different relationship types. The authors introduce GATNE, a framework for both transductive (GATNE-T) and inductive (GATNE-I) embedding learning. GATNE integrates node attributes and leverages multi-view relationships characteristic of AMHENs. In this paradigm, nodes are mapped to low-dimensional vectors while preserving network properties. The GATNE model splits each node's embedding into a base embedding shared across edge types, and edge-specific embeddings aggregated from neighbors. GATNE-T is the transductive model using random walks and heterogeneous skip-gram on the network. GATNE-I is the inductive model using node attributes to generate base and edge embeddings via transformation functions.

\subsection{Knowledge Graph Embeddings}
The use of Knowledge Graphs (KG) \citep{wang2017knowledge} has been gaining traction in representing diverse types of information through various entities and relations. The KG is a directed heterogeneous multigraph where node and relation types provide domain-specific semantics. This structure allows for encoding knowledge in a manner interpretable by humans and conducive to automated analysis. In the context of KGs, the conventional graph terminology is replaced with terms like entities for vertices and triplets for directed edges. The triplet is represented as a (h, r, t) tuple in which 'h' is the head entity, 't' is the tail entity, and 'r' is the relation linking them. The term 'relation' here specifies the type of relation, such as 'wants-to-buy' or 'has-bought'. Previous work has shown that a populated KG effectively encodes the knowledge about a particular domain based on the types of nodes and relations included. \cite{heiko}

\subsection{Exploiting Reviews for recommender models}
User reviews provide rich insights into user experiences, preferences, and opinions. Deep Cooperative Neural Networks (DeepCoNN) \citep{zheng2017joint} learn item properties and user behaviors jointly from review text to potentially improve the quality of recommendations and alleviate the sparsity problem. They further enhance the model interpretability by exploiting rich semantic representations of review texts with CNNs. In their work, one of the CNN networks models user behavior using the reviews written by the user, and the other network models item properties using the written reviews for the item. Other works like \citet{wu2017joint} proposes a co-trained neural network with recurrent architecture for combining numerical ratings, natural language reviews, and temporal dynamics to achieve highly accurate recommendations. \citet{tay2018multi} proposed a review based recommendation system with multi-pointer co-attention.  

\vspace{0.2in}
In this work, we utilize item reviews to extract positive aspects of items and harness the encoding power of Knowledge Graphs to generate features for our items by linking aspects and other meta-information to train the GATNE model for generating relevant recommendations for anchor items along with their header texts. We highlight our methodology in detail in the next section. 

\section{Methodology}
Our methodology revolves around training a Graph Neural Network (GNN) model to generate embedding representations for items, which are then utilized to generate relevant items for a given set of anchor items. The following subsections present details of each component of the proposed DTS model.

\subsection{Data Collection}  
E-commerce session data captures the user's journey through the website within a single browsing period. Analyzing co-bought and co-add-to-cart patterns within  session data can reveal complementary item relationships. This strategy helps identify frequently purchased together items, potentially uncovering implicit purchase intentions and informing product bundling strategies. We restricted our data for a time frame of 3 months and top-5 performing product categories, as shown in  \ref{table:2}. Within the product hierarchy, each broad product category is further subdivided into more granular classifications known as product types. For instance, the \textit{Home \& Garden} product category encompasses a variety of product types, including Garden Center, Grills \& Outdoor Cooking, Indoor \& Outdoor Plants, Outdoor Power Equipment, Pool Supplies, Patio Furniture, Flooring, Hardware, Kitchen Renovation, Paint, Tools, Wall Coverings etc.  We construct the network $G$, where all item nodes are represented as $\mathcal{V}$, and edge types- \textit{co-bought} and \textit{co-add-to-cart} are represented as $\mathcal{E}$. We consider edges between items only if both items are from the same product category. If both items are bought in the session, they are connected by co-bought edge, if the items are added-to-cart in the same session, they are connected by co-add-to-cart edge.   
Our training data consists of two overlapping sets: 1) items interacted by customers on the e-commerce website, including items added to cart and transacted within the chosen time period, denoted as  $\mathcal{V}_{a}$ and 2) items with extracted aspects from their reviews available on the website, denoted as  $\mathcal{V}_{b}$. As these items possess a sufficient number of reviews suitable for aspect extraction, it is reasonable to categorize them as popular items. The anchor set, a subset of $\mathcal{V}_{a}$, consists of high performing items for which we generate a DTS recommendation carousel. The items of set $\mathcal{V}_{b}$ form the recall set of items. Our anchor set consists of 7 million nodes while the recall set consists of 1.1 million nodes. In the hierarchical classification of our product catalogue, the 1.1 million nodes with aspects are distributed according to Table \ref{table:1}.

\begin{table}[ht]
\caption{Summary of the product catalogue} 
\vspace{5mm}
\centering
\begin{tabular}{ll@{\hspace{3mm}}ll} 
\toprule
Product Category & Number of items & Number of ptypes\\
\toprule
Home \& Garden & 223,395  & 1539\\
Clothing, Shoes \& Accessories & 140,861 & 244\\
Books, Music \& Movies & 126,522 & 12\\
Health \& Beauty  & 107,009 & 583\\
Electronics & 49,316 & 346\\
\bottomrule
\end{tabular}
\label{table:1}
\end{table}

\begin{table*}[h]
\caption{Sample extracted aspects and corresponding headers for our product catalogue.}
\vspace{5mm}
\centering
\begin{tabularx}{\textwidth}{|>{\centering\arraybackslash}X|X|X|} 
\hline
Item & Extracted Aspects & Headers \\
\hline
\raisebox{\dimexpr-\height+\ht\strutbox\relax}{\includegraphics[width=2cm,height=3cm,keepaspectratio]{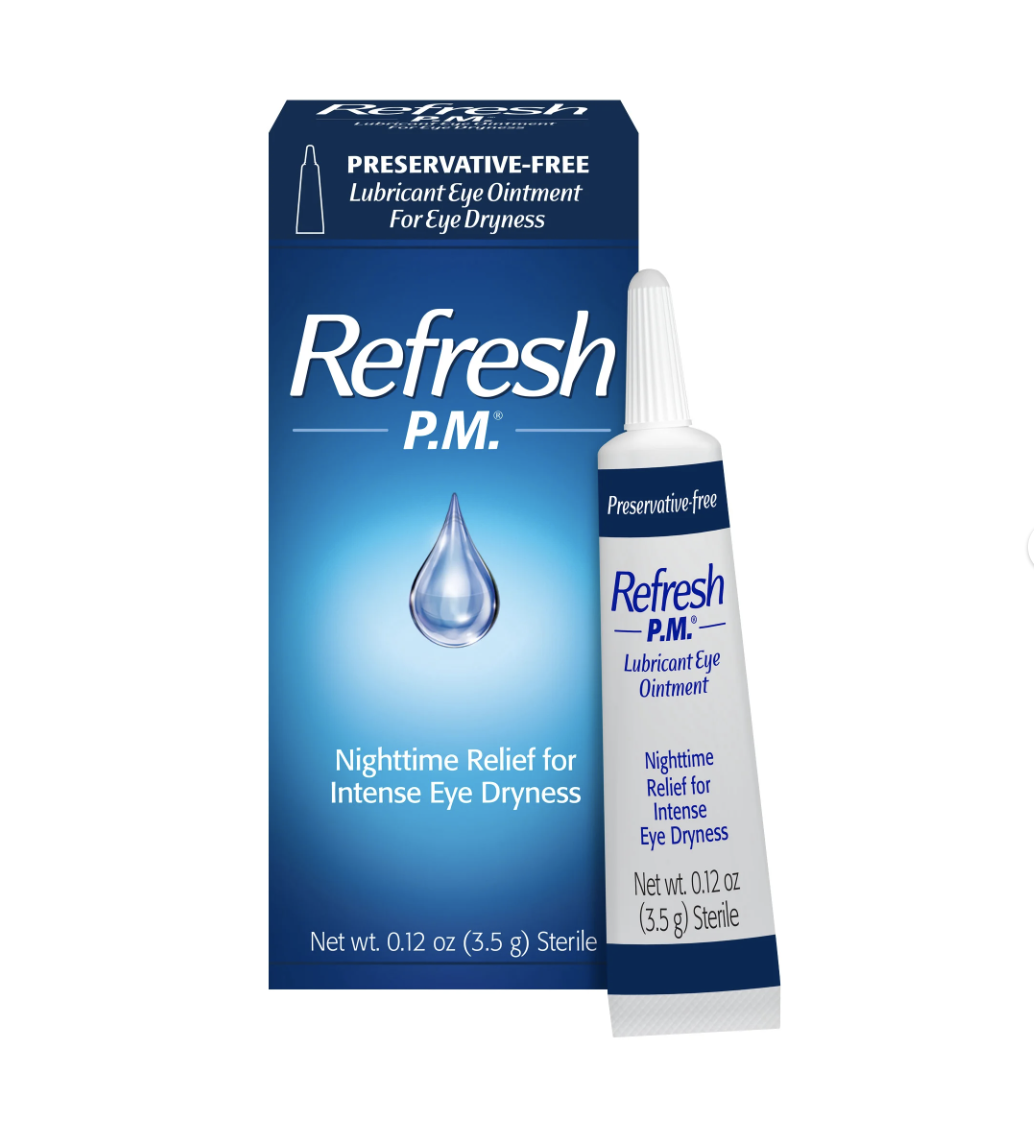}} &
\begin{itemize}[leftmargin=*]
  \item best gel
  \item great for dry eyes
  \item excellent ointment
\end{itemize} & 
\begin{itemize}[leftmargin=*]
  \item Customers say these are best gel
  \item Similar items that are great for dry eyes
  \item Customers say these are excellent ointment
\end{itemize} \\
\hline 
\raisebox{\dimexpr-\height+\ht\strutbox\relax}{\includegraphics[width=2cm,height=3cm,keepaspectratio]{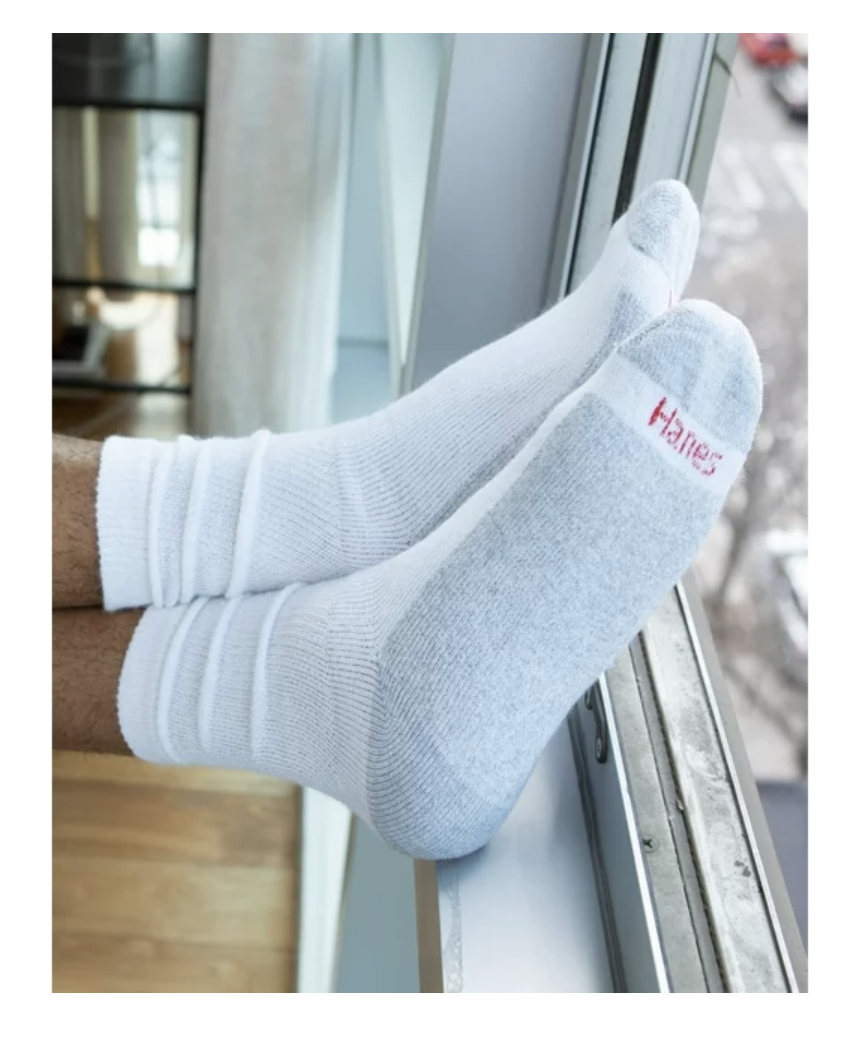}} & 
\begin{itemize}[leftmargin=*]
  \item great fit
  \item comfortable
  \item great cushioning
  \item wear well
\end{itemize} & 
\begin{itemize}[leftmargin=*]
  \item Customers say these are a great fit
  \item Customers say these are comfortable
  \item Customers say these have great cushioning
  \item Similar items that wear well
\end{itemize} \\
\hline
\end{tabularx}
\label{table:2}
\end{table*}

\subsection{Items' Aspects Extraction from User Reviews}
\label{sec3_2}
Our proprietary ensemble linguistic model extracts these aspects and corresponding headers for each item from user reviews and assigns a relevance score to each aspect. This model primarily uses BERT-based aspect extraction, aspect-sentiment classification and aggregation. The goal of the aggregation step in the pipeline is to draw category-level and product-level insights based on all the individual opinions extracted within a category. In the aggregation, the following steps happen: Aspect unification (unifying different forms of aspects with the same meaning), 
Statistical validation (applying filters to remove outliers), 
Aspect naming (determining how to represent each aspect), 
Ontology creation (identifying the aspects hierarchy), 
Scoring (computing aspect scores per product). Aspect extraction is formulated as an iterative process. This approach leverages human intervention for labeling data and subsequent retraining of the aspect extraction BERT-based model.
The 1.1 million nodes with aspects capture a total of 8354 distinct aspects, represented as $\mathcal{A}$. These aspects vary from- \textit{great characters}, \textit{great story} to \textit{stylish} and \textit{love the feel} to provide few examples. These aspects map to two header syntactic variations each starting with \textit{Similar items} or \textit{Customers say}.  For example, \textit{Similar items that are stylish} and  \textit{Customers say these are stylish} are two possible headers for the aspect - \textit{stylish}. These headers are generated automatically, however evaluated manually for lexical clarity, conciseness, proper grammar, and overall readability. Finally, we have $N_i$ extracted \{aspects, relevance score\} pairs as $\{a_{n_i}, {asp\_rel}_{n_i}\}$, where $(1\leq n_i \leq N_i)$ for $v_i$ item. $\mathcal{A}_{v_i}$ denotes the set of aspects for item $v_i$, i.e. $a_{n_i} \in \mathcal{A}_{v_i}$. Table \ref{table:2} presents sample example of the extracted aspects and headers.
Table \ref{table:3} presents a summary of number of aspects per item. On an average, each item has 8 extracted aspects and corresponding headers. 


\begin{table}[ht]
\caption{Summary of number of aspects per item}
\vspace{5mm}
\centering
\begin{tabular}{ll@{\hspace{8mm}}ll} 
\toprule
mean & stddev & min & max \\
\toprule
8.61 & 14.95 & 1 & 419 \\
\bottomrule
\end{tabular}
\label{table:3}
\end{table}


\subsection{E-commerce Graph Creation}
\label{sec_graph}
\begin{figure}[ht]
\centering
\includegraphics[width=\columnwidth]{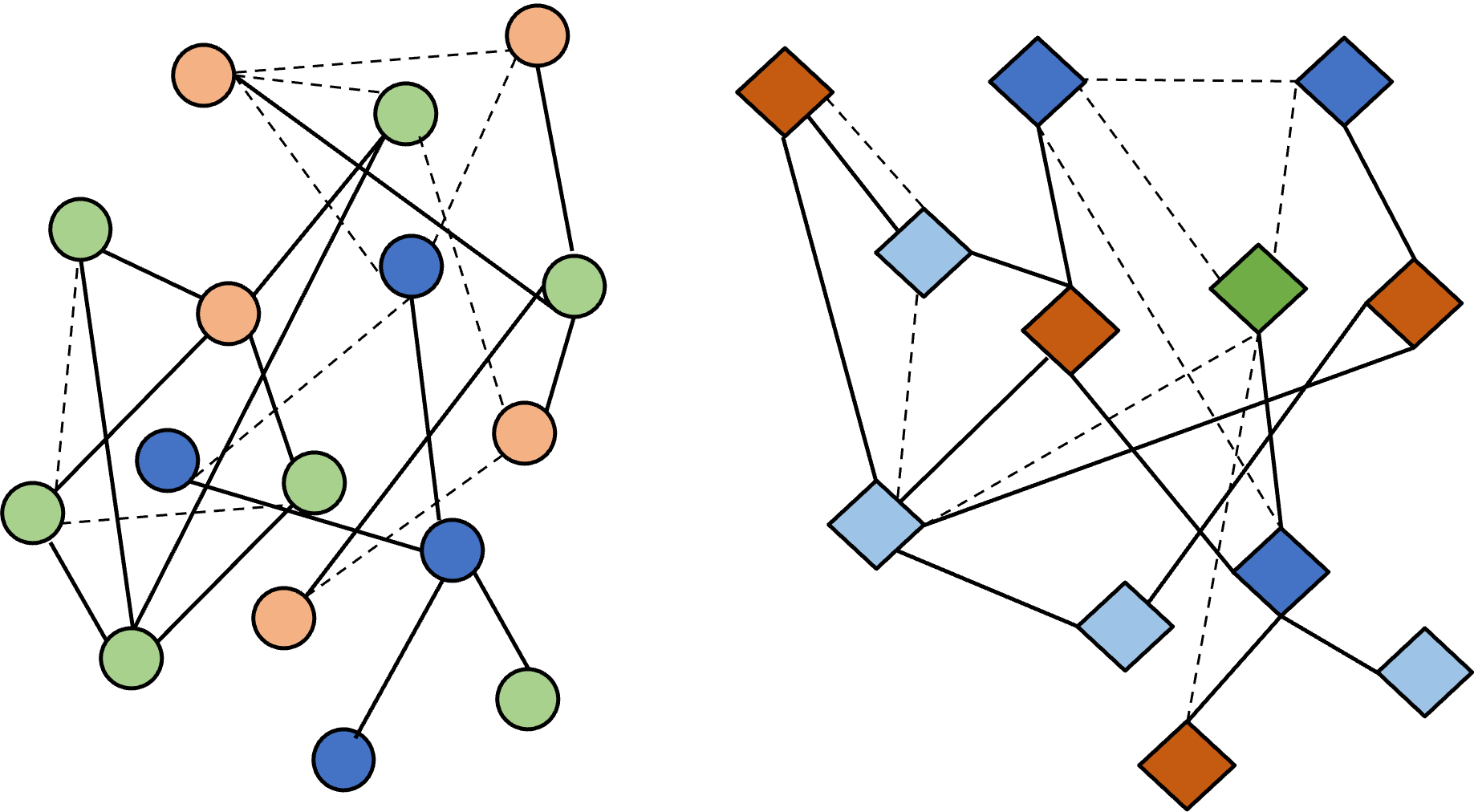}
    \caption{Sample e-commerce graph, where item nodes are connected by \textit{co-add-to-cart} (denoted by dashed lines) and \textit{co-bought} (denoted by solid lines). Each sub-graph corresponds to each product category as edges are restricted among items of the same product type.}
    \vspace{7mm}
    \label{fig:fig3}
\end{figure}

The items collected from the two overlapping sets, $\mathcal{V}_{a}$ and $\mathcal{V}_{b}$ are combined to construct a graph where nodes represent items, and edges represent customer interactions.  Figure \ref{fig:fig3} shows a simplified e-commerce items' graph, where the item node shape represents the product category while the item node color represents the granular product type. Since edges are restricted to items from the same product category. $G$ comprises of several sub-graphs corresponding to each product category. Each sub-graph has two edge types - \textit{co-add-to-cart} edge type denotes adding items to the shopping cart while \textit{co-bought} edge type denotes completing transactions.

\subsection{Item Base Embedding} 
\label{sec_base_emb}
In training the GATNE model, two primary approaches exist for initializing node embeddings: random initialization and attribute-derived embeddings. Given our focus on aspects within the DTS model, we opt for attribute-derived base embeddings. This selection allows the model to leverage the inherent meaning of item attributes during the embedding process. For the purpose of learning base embedding representations, we constructed a separate Knowledge Graph (KG) using the e-commerce item catalogue information and an existing item similarity model. Table ~\ref{table:4} details the different tuples encoded for the knowledge graph training stage. These tuples leverage information extracted from various sources:
\begin{itemize}
  \item \textbf{Item Catalog Data}: \textit{Product type, Price Band}, and \textit{Brand} relations are directly derived from the item catalog
  \item \textbf{Item Aspects Data}: The \textit{Aspect} relation is established based on information described in Section ~\ref{sec3_2}
  \item \textbf{Proprietary Model}: The "Similarity" relation is constructed using our proprietary item-to-item similarity model.
\end{itemize}

\begin{table}[h]
\caption{Summary of the product catalogue}
\vspace{5mm}
\centering
\begin{tabular}{ll@{\hspace{8mm}}ll} 
\toprule
Head Entity & Relation & Tail Entity \\
\toprule
Item & \textit{Product type} & Aspect   \\
Item & \textit{Aspect} & Carousel Title   \\
Item & \textit{Similarity} & Item   \\
Item & \textit{Price Band} & Price Band Value (0-5)   \\
Item & \textit{Brand} & Brand popularity rank   \\
\bottomrule
\end{tabular}
\label{table:4}
\end{table}

\noindent DGL-KE (Deep Graph Library Knowledge Embedding) \citep{DGL-KE}  is a framework specifically designed for representing and reasoning over knowledge graphs within the Deep Graph Library (DGL). It facilitates the creation of knowledge graph embeddings by translating entities and relations into low-dimensional vector representations. This enables tasks like link prediction, entity classification, and knowledge graph completion within the DGL ecosystem.
We leverage DGL-KE DistMult\cite{yang2015embedding} to generate item base embeddings for our GATNE model. Our implementation closely follows the open source repository available on DGL website \citep{dglke_website}. This approach allows us to encode rich relational information from the knowledge graph into low-dimensional vector representations of items, denoted as ${x}_i \in \mathbb{R}^{d_b}$. These learned item embeddings are then directly fed into the GATNE model as node \textbf{base embedding} for further processing and utilization within the network architecture.

\subsection{Training GATNE model}

We train the GATNE \citep{cen2019representation} model on the e-commerce graph, $G$, defined in section-\ref{sec_graph}. The overall item embedding for each edge type comprises of two parts: \textbf{shared embedding} and \textbf{edge embedding}. The shared embedding of node $v_i$ is shared between different edge types. 

\noindent In our graph representation, the edge embedding $\mathbf{u}^{(k)}_{i,r} \in \mathbb{R}^s$ for node $v_i$ of relation type $r$ at the $k$-th level $(1\leq k\leq K)$ aggregates the neighbors' edge embeddings using a mean aggregator \citep{xu2018powerful} as: 
\begin{equation}
\label{mean}
  \mathbf{u}^{(k)}_{i,r} = \sigma(\mathbf{\hat{W}}^{(k)} \cdot \operatorname{mean}(\{\mathbf{u}^{(k-1)}_{j,r}, \forall v_j\in\mathcal{N}_{i,r}\})),
\end{equation}

 \noindent where $W$ is a learnable linear mapping matrix, $\sigma$ is an activation function,  $s$ is the dimension of edge embeddings, $K$ is the total aggregation levels and $\mathcal{N}_{i,r}$ contains the neighbors of node $v_i$ on edge type $r$. 
 The initial edge embeddings $\mathbf{u}^{(0)}_{i,r}$ for node $i$ on edge type $r$ is parameterized as the function of base embedding $\mathbf{x}_i$ as $\mathbf{u}^{(0)}_{i,r}=\mathbf{g}_{r}(\mathbf{x}_i)$, where $\mathbf{g}_{r}$ is a transformation function that transforms the base embedding to an edge embedding of node $v_i$ on the edge type $r$. Following the aggregation process, we denote the final, $K$-th level edge embedding $\mathbf{u}^{(K)}_{i,r}$ for node $v_i$ on relation type $r$ as edge embedding $\mathbf{u}_{i,r}$. Subsequently, we concatenate all the edge embeddings associated with node $v_i$ as $\mathbf{U}_i$ with size $s$-by-$m$, where $s$ represents the individual edge embedding dimension and $m$ represents the total number of relation types considered.
\begin{equation}
\mathbf{U}_i = ( \mathbf{u}_{i,1}, \mathbf{u}_{i,2}).
\end{equation}

\noindent To effectively integrate information from all relevant edge types for a specific node, we leverage a self-attention mechanism \cite{lin2017structured} to get the coefficients $\mathbf{a}_{i,r} \in \mathbb{R}^m$ to compute  of linear combination of vectors in $\mathbf{U}_i$ on edge type $r$ as:
\begin{equation}
\label{eqn:air}
\mathbf{a}_{i,r} = \operatorname{softmax} ( \mathbf{w}_r^{T} \tanh ~( \mathbf{W}_r \mathbf{U}_i ) )^T,
\end{equation}
\noindent where $\mathbf{w}_r$ and $\mathbf{W}_r$ are trainable parameters for edge type $r$ with size $d_a$ and $d_a\times s$ respectively.

\noindent Overall Node embedding for node $\mathbf{v}_{i,r}$ for a given relation type $r$ is computed as: 

\begin{equation}
\label{eqn:1}
\mathbf{v}_{i,r} = \mathbf{h}(\mathbf{x}_i) + \alpha_r \mathbf{M}_r^{T} \mathbf{U}_i \mathbf{a}_{i,r} + \beta_r \mathbf{D}^T \mathbf{x}_i,
\end{equation}
where ${x}_i$ represents the base item embedding learned from the DGL-KE model, $\mathbf{h}$ is a multi-layer perceptron \citep{pal1992multilayer} transformation function that transforms $\mathbf{x}_i$ to shared embedding. $\alpha_r,\beta_r$ are the hyper-parameter denoting the importance of shared embedding and edge embeddings respectively towards the overall embedding, set as 0.5 each based on empirical hyper-parameter tuning. $\mathbf{D} \in \mathbb{R}^{d_b \times d}$, $\mathbf{M}_r \in \mathbb{R}^{s\times d}$  are the trainable transformation matrices. \\

\noindent Our model optimization leverages random walks to generate sequences of connected nodes within the graph. These node sequences are then employed as input for a skip-gram model \citep{mikolov2013efficient, mikolov2013distributed}, which facilitates the learning of low-dimensional embeddings for the item nodes. We use negative log-likelihood and negative sampling to approximate the objective function as defined in \cite{cen2019representation}.

\begin{algorithm}[ht]
\KwData{co-bought and co-add-to-cart items from e-commerce sessions data,  item attributes data}
\KwIn{e-commerce graph $G=(\mathcal{V},\mathcal{E},\mathcal{A})$, embedding dimension $d$, random walk length $l$, learning rate $\eta$, negative samples $N$, coefficient $\alpha$, $\beta$. }
\KwOut{overall embedding $\mathbf{v}_{i,r}$ for all nodes for edge type $r \in $ \{co-bought, co-add-to-cart\}}


Model parameters ($\theta$) initialization.\\
Base item embedding initialization as generated in section-\ref{sec_base_emb}
Random walks generation: $\mathcal{W}_r$ for each edge type $r$.\\
Training samples generation:  $\{(v_i,v_j,r)\}$ from random walks $\mathcal{W}_r$ for each edge type $r$.
 
\While{not converged} {
    \ForEach{$(v_i,v_j,r)\in$ training samples} {
        Compute $\mathbf{v}_{i,r}$ using Equation (\ref{eqn:1})\\
        Sample $L$ negative samples and calculate skip-gram negative log-likelihood objective function.\\
        Update model parameters $\theta$.
    }
}


 \caption{Dynamic Text Snippets (DTS) Training}
\end{algorithm}

\subsection{Recall generation}

To generate recommendations, an Approximate Nearest Neighbor (ANN) index is employed to retrieve the nearest neighbors for a given set of anchor items along with relevance score using FAISS \citep{douze2024faiss, johnson2019billion} \textit{IndexFlatL2} index. For FAISS search, we restrict the recall set to items within $\mathcal{V}_{b}$. This strategic selection is motivated by the availability of extracted aspects for these items. By focusing on this subset, DTS recommendation carousel groups items that share same aspects, explained by the corresponding header. For each anchor item $v_i$, we retrieve 50 ANN recall set, denoted as $\mathcal{R}_i$ consisting of recall items $\{(r^p_i, {faiss\_rel}^p_i),  (1\leq p\leq 50)\}$, where ${faiss\_rel}^p_i$ is the FAISS relevance score for the corresponding recall item $r^p_i$. $\mathcal{A}_i$ is the set of unique aspects in the recall set $\mathcal{R}_i$.

\subsection{Ranking, Aspect selection and Header Selection} 

With the anchor items and the corresponding recommendation set obtained from the ANN index, the most frequently occurring titles among the recommended items are extracted. These titles are then mapped to the recommendation list, creating a one-to-many mapping structure suitable for displaying recommendations in a carousel format on the website. Each anchor item $v_i$ is associated with the recall set $\mathcal{R}_i$. $\mathcal{A}_{r^p_i}$ is the set of unique aspects for the recall item $r^p_i$. Header score for a given recall item $r^p_i$ and aspect $a$ pair is given as ${faiss\_rel}(r^p_i, v_i) \times {asp\_rel}(a, {r^p_i})$. $N_a$ represents the number of items returned in the recall set that share aspect $a$. Header score for a given aspect  is computed as the average of the individual header scores assigned to all recall items having that aspect. 

\begin{equation}
\label{eqn:title}
\mathbf{{asp}_{sel}} =  \arg \max_{a \in \mathcal{A}_i} \biggl( \frac{1}{N_a}\sum_{\substack{r^p_i \in \mathcal{R}_i \\ a \in \mathcal{A}_{r^p_i}}} {faiss\_rel}(r^p_i, v_i) \times {asp\_rel}(a, {r^p_i}) \biggl)
\end{equation}
Finally, the header with the highest aspect header score is chosen for DTS carousel as shown in equation-\ref{eqn:title}. \\

\begin{figure}[ht]
\centering
\includegraphics[width=\columnwidth]{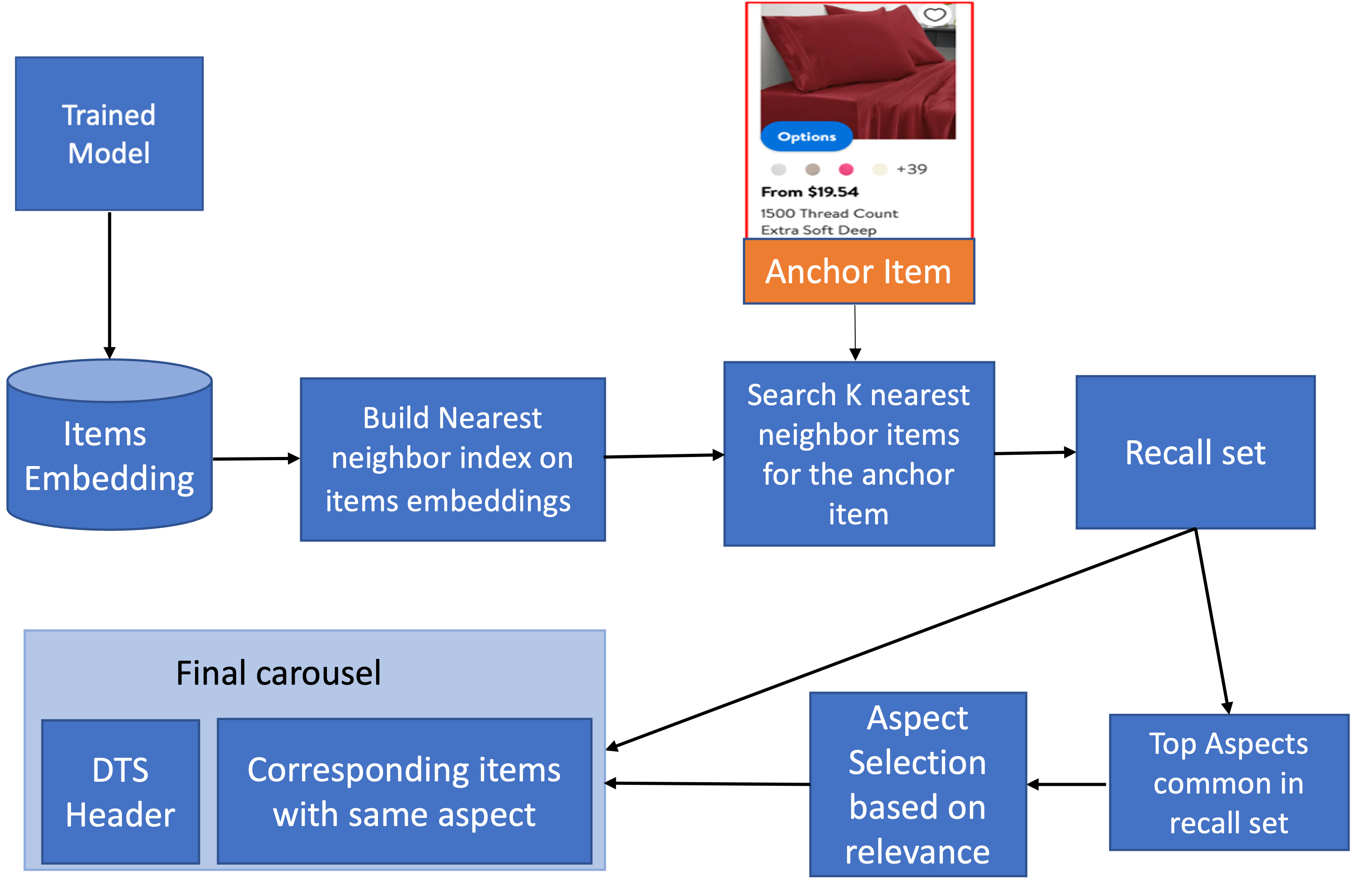}
    \caption{Dynamic Text Snippets (DTS) Inference Pipeline}
    \label{fig:inference}
\end{figure}

Figure-\ref{fig:inference} shows the overall inference pipeline for the DTS model, where the final carousel shows the most relevant header along with corresponding recall items possessing that aspect. 

\section{Experiments and Results}

\subsection{Training Parameters}  

We utilized the DGL-KE open-source package \cite{DGL-KE} for training the DistMult model. The parameters chosen for the base embedding generation model are presented in Table \ref{table:6}.

\begin{table}[ht]
\caption{DistMult Training Parameters}
\vspace{5mm}
\centering
\begin{tabular}{ll@{\hspace{8mm}}ll} 
\toprule
Parameter & Value  \\
\toprule
batch size & 1000   \\
neg\_sample\_size & 200   \\
regularization\_coef & 1e-9   \\
hidden\_dim & 128   \\
gamma & 19.9   \\
lr & 0.25   \\
\bottomrule
\end{tabular}
\label{table:6}
\end{table}

\begin{table}[ht]
\caption{GATNE Training Parameters}
\vspace{5mm}
\centering
\begin{tabular}{ll@{\hspace{8mm}}ll} 
\toprule
Parameter & Value  \\
\toprule
embedding\_size ($d$, $d_b$) & 128   \\
embedding\_u\_size ($s$) & 10   \\
edge\_type\_count ($|\mathcal{E}|$) & 2   \\
negative samples ($N$) & 4  \\
neighbour\_samples (for weight update) & [5] \\
dim\_a ($d_a$) & 20   \\
size\_of\_window (for generating training pairs) & 3 \\
random\_walk\_length ($l$) & 5 \\
num\_walks & 5 \\
lr & 1e-3 \\
optimizer & Adam \\
\bottomrule
\end{tabular}
\label{table:7}
\end{table}

\noindent The GATNE model implementation closely resembles the DGL open source implementation \citep{wang2019dgl}. To scale the graph for incorporating millions of nodes in the training process, we generate the random walks and training pairs separately and utilize these during training instead of creating walks every training iteration. The parameters for the best performing GATNE model are presented in Table \ref{table:7}.

\subsection{Base Item Embedding Ablation Result} 
To comprehensively evaluate base item embedding representation strategies, we investigated the use of BERT-based embeddings and compared their performance with DGL-KE as inputs to the GATNE model. As \textit{Home and Garden} is the product category with maximum items that contain aspects, we created a subset of 30K items for experimenting with different base embedding representations. We used these base embedding variations, $x_i$ to train the GATNE model, create DTS recall set and compare its performance against our website's best performing recommendation model. The results are reported in Table \ref{table:5}. 

\begin{table}[ht]
\caption{Performance of different base embedding representations}
\vspace{5mm}
\centering
\begin{tabular}{ll@{\hspace{8mm}}ll} 
\toprule
Model & Overall NDCG Score \\
\toprule
DistMult KE & 0.50   \\
Tiny Bert \cite{reimers-2019-sentence-bert}   & 0.52 \\
\textsuperscript{(sentence-transformers/paraphrase-TinyBERT-L6-v2)} & \\
\bottomrule
\end{tabular}
\label{table:5}
\end{table}

While linguistic BERT-based features were considered during our model prototyping, they were ultimately not chosen due to challenges in scaling the 768-dimension feature representations during model deployment on the available resources. 

The NDCG score affirmed the relevance of our recommendations, however, due to the lack of a baseline model for comparison due to the novelty of our design, we decided to subject our model to an A/B test utilizing live traffic in order to ascertain its robustness and effectiveness. 

\subsection{A/B Test Result} 

We followed a standard GNN training paradigm in which we observed a minimization and saturation of loss values during the embedding generation step for the graph with 7 million nodes. Following this, our recall set was generated along with multiple titles for each anchor set.  

We then initiated an A/B test involving millions of customers to evaluate the functionality of our model in a real-world setting. Initially, we limited traffic exposure to 5\%, but upon observing satisfactory metrics, we expanded exposure to full traffic across all platforms, including Web, iOS, and Android. Our test was conducted for traffic on web pages of anchor items for which we had recommendations for the DTS carousel. While the control and variation shared two common carousels, there was a notable distinction in the third carousel, with the variation containing our DTS model as a recommendation carousel. 

The A/B test lasted for two weeks with 100\% traffic, during which time our model consistently demonstrated positive performance. We observed a GMV lift of 0.05\% on Web, 0.07\% on Android and 0.04\% on iOS. As a result, it was subsequently launched on the website. This suggests that despite the lack of a conventional baseline model for comparison, our novel model has proven its effectiveness in a live testing environment, thereby validating its utility and potential for wider application.

While delving deeper into the results, we observed a notable consistency across all platforms where the overall add-to-cart (ATC) feature remained neutral. However, there was an observable net increase in ATC and item clicks when calculated for the combination of these three carousels. This potentially indicated that users were locating relevant information more quickly and efficiently.

This outcome was uniform for Android, Web, and iOS platforms. When we dissected the data platform-wise, we observed that for Android, the highest net ATC changes were recorded in Media \& Gaming, Electronics, and Sporting goods. However, Grocery and personal care exhibited a negative change which may suggest a need for improvement in these areas.

For both the Web and iOS platforms, Electronics, Grocery, and Home management were found to contribute significantly to the overall increase in ATC. These results highlight the nuances of user behavior and preferences across different platforms, and suggest targeted strategies might be necessary to optimize user engagement.

Owing to its overall positive performance, the DTS recommendation generation pipeline is successfully deployed and integrated into the live website, enabling dynamic headers for item recommendations captured by the graph representation. 

\section{Conclusion and Future Work} 

This study has highlighted the potential of utilizing user-generated reviews as a tool for creating recommender systems with dynamic text snippets (DTS). Our proposed model presents an encouraging method for improving the recommendation experience by integrating the aspects that users correlate with items. This approach can significantly contribute to more informed decision-making processes and enhance user satisfaction.

Our work has provided a solid foundation for future studies in this field. With a baseline model established, subsequent work will endeavor to incorporate other modelling techniques to improve the performance of aspect-based recommender systems further. One promising direction is the transition from dynamic text snippets (DTS) to personalized text snippets (PTS). This will entail mapping user affinities to aspects, thereby enabling us to recommend not only items but also header texts directly to users during their browsing sessions. While Generative AI offers exciting possibilities for dynamic header generation, the grounding problem remains a significant hurdle. This limitation arises because, in our context, the recommended items must have the chosen aspect for the generated header to be truly informative. Overcoming this grounding problem by incorporate item catalog knowledge graph and reviews reasoning capabilities should be another promising direction.

Through this strategy, we hope to create a more personalized and engaging user experience. This could potentially improve the way recommender systems function, making them even more responsive and tailored to individual user needs and preferences. We believe that our work paves way for exciting advancements in the field of recommendation systems and offers substantial opportunities for future contributions.

\bibliography{p86}

\begin{thebibliography}{25}
\providecommand{\natexlab}[1]{#1}
\providecommand{\url}[1]{\texttt{#1}}
\expandafter\ifx\csname urlstyle\endcsname\relax
  \providecommand{\doi}[1]{doi: #1}\else
  \providecommand{\doi}{doi: \begingroup \urlstyle{rm}\Url}\fi

\bibitem[Cen et~al.(2019)Cen, Zou, Zhang, Yang, Zhou, and Tang]{cen2019representation}
Y.~Cen, X.~Zou, J.~Zhang, H.~Yang, J.~Zhou, and J.~Tang.
\newblock Representation learning for attributed multiplex heterogeneous network.
\newblock In \emph{Proceedings of the 25th ACM SIGKDD international conference on knowledge discovery \& data mining}, pages 1358--1368, 2019.

\bibitem[Douze et~al.(2024)Douze, Guzhva, Deng, Johnson, Szilvasy, Mazaré, Lomeli, Hosseini, and Jégou]{douze2024faiss}
M.~Douze, A.~Guzhva, C.~Deng, J.~Johnson, G.~Szilvasy, P.-E. Mazaré, M.~Lomeli, L.~Hosseini, and H.~Jégou.
\newblock The faiss library.
\newblock 2024.

\bibitem[Jannach and Jugovac(2019)]{jannach2019measuring}
D.~Jannach and M.~Jugovac.
\newblock Measuring the business value of recommender systems.
\newblock \emph{ACM Transactions on Management Information Systems (TMIS)}, 10\penalty0 (4):\penalty0 1--23, 2019.

\bibitem[Johnson et~al.(2019)Johnson, Douze, and J{\'e}gou]{johnson2019billion}
J.~Johnson, M.~Douze, and H.~J{\'e}gou.
\newblock Billion-scale similarity search with {GPUs}.
\newblock \emph{IEEE Transactions on Big Data}, 7\penalty0 (3):\penalty0 535--547, 2019.

\bibitem[Konstan and Riedl(2012)]{konstan2012recommender}
J.~A. Konstan and J.~Riedl.
\newblock Recommender systems: from algorithms to user experience.
\newblock \emph{User modeling and user-adapted interaction}, 22:\penalty0 101--123, 2012.

\bibitem[Labs(2020)]{dglke_website}
A.~Labs.
\newblock {DGL-KE} documentation, 2020.
\newblock URL \url{dglke.dgl.ai/doc/}.

\bibitem[Lin et~al.(2017)Lin, Feng, Santos, Yu, Xiang, Zhou, and Bengio]{lin2017structured}
Z.~Lin, M.~Feng, C.~N.~d. Santos, M.~Yu, B.~Xiang, B.~Zhou, and Y.~Bengio.
\newblock A structured self-attentive sentence embedding.
\newblock \emph{arXiv preprint arXiv:1703.03130}, 2017.

\bibitem[McNee et~al.(2006)McNee, Riedl, and Konstan]{mcnee2006being}
S.~M. McNee, J.~Riedl, and J.~A. Konstan.
\newblock Being accurate is not enough: how accuracy metrics have hurt recommender systems.
\newblock In \emph{CHI'06 extended abstracts on Human factors in computing systems}, pages 1097--1101, 2006.

\bibitem[Mikolov et~al.(2013{\natexlab{a}})Mikolov, Chen, Corrado, and Dean]{mikolov2013efficient}
T.~Mikolov, K.~Chen, G.~Corrado, and J.~Dean.
\newblock Efficient estimation of word representations in vector space.
\newblock \emph{arXiv preprint arXiv:1301.3781}, 2013{\natexlab{a}}.

\bibitem[Mikolov et~al.(2013{\natexlab{b}})Mikolov, Sutskever, Chen, Corrado, and Dean]{mikolov2013distributed}
T.~Mikolov, I.~Sutskever, K.~Chen, G.~S. Corrado, and J.~Dean.
\newblock Distributed representations of words and phrases and their compositionality.
\newblock \emph{Advances in neural information processing systems}, 26, 2013{\natexlab{b}}.

\bibitem[Pal and Mitra(1992)]{pal1992multilayer}
S.~K. Pal and S.~Mitra.
\newblock Multilayer perceptron, fuzzy sets, classifiaction.
\newblock 1992.

\bibitem[Paulheim(2016)]{heiko}
H.~Paulheim.
\newblock Knowledge graph refinement: A survey of approaches and evaluation methods, 2016.
\newblock URL \url{https://semantic-web-journal.net/system/files/swj1167.pdf}.

\bibitem[Reimers and Gurevych(2019)]{reimers-2019-sentence-bert}
N.~Reimers and I.~Gurevych.
\newblock Sentence-bert: Sentence embeddings using siamese bert-networks.
\newblock In \emph{Proceedings of the 2019 Conference on Empirical Methods in Natural Language Processing}. Association for Computational Linguistics, 11 2019.
\newblock URL \url{http://arxiv.org/abs/1908.10084}.

\bibitem[Tarnowska et~al.(2020)Tarnowska, Ras, and Daniel]{tarnowska2020recommender}
K.~Tarnowska, Z.~W. Ras, and L.~Daniel.
\newblock \emph{Recommender system for improving customer loyalty}, volume~1.
\newblock Springer, 2020.

\bibitem[Tay et~al.(2018)Tay, Luu, and Hui]{tay2018multi}
Y.~Tay, A.~T. Luu, and S.~C. Hui.
\newblock Multi-pointer co-attention networks for recommendation.
\newblock In \emph{Proceedings of the 24th ACM SIGKDD international conference on knowledge discovery \& data mining}, pages 2309--2318, 2018.

\bibitem[Veli{\v{c}}kovi{\'c} et~al.(2017)Veli{\v{c}}kovi{\'c}, Cucurull, Casanova, Romero, Lio, and Bengio]{velivckovic2017graph}
P.~Veli{\v{c}}kovi{\'c}, G.~Cucurull, A.~Casanova, A.~Romero, P.~Lio, and Y.~Bengio.
\newblock Graph attention networks.
\newblock \emph{arXiv preprint arXiv:1710.10903}, 2017.

\bibitem[Wang et~al.(2019)Wang, Zheng, Ye, Gan, Li, Song, Zhou, Ma, Yu, Gai, Xiao, He, Karypis, Li, and Zhang]{wang2019dgl}
M.~Wang, D.~Zheng, Z.~Ye, Q.~Gan, M.~Li, X.~Song, J.~Zhou, C.~Ma, L.~Yu, Y.~Gai, T.~Xiao, T.~He, G.~Karypis, J.~Li, and Z.~Zhang.
\newblock Deep graph library: A graph-centric, highly-performant package for graph neural networks.
\newblock \emph{arXiv preprint arXiv:1909.01315}, 2019.

\bibitem[Wang et~al.(2017)Wang, Mao, Wang, and Guo]{wang2017knowledge}
Q.~Wang, Z.~Mao, B.~Wang, and L.~Guo.
\newblock Knowledge graph embedding: A survey of approaches and applications.
\newblock \emph{IEEE transactions on knowledge and data engineering}, 29\penalty0 (12):\penalty0 2724--2743, 2017.

\bibitem[Wu et~al.(2017)Wu, Ahmed, Beutel, and Smola]{wu2017joint}
C.-Y. Wu, A.~Ahmed, A.~Beutel, and A.~J. Smola.
\newblock Joint training of ratings and reviews with recurrent recommender networks.
\newblock 2017.

\bibitem[Wu et~al.(2022)Wu, Sun, Zhang, Xie, and Cui]{10.1145/3535101}
S.~Wu, F.~Sun, W.~Zhang, X.~Xie, and B.~Cui.
\newblock Graph neural networks in recommender systems: A survey.
\newblock \emph{ACM Comput. Surv.}, 55\penalty0 (5), dec 2022.
\newblock ISSN 0360-0300.
\newblock \doi{10.1145/3535101}.
\newblock URL \url{https://doi.org/10.1145/3535101}.

\bibitem[Xu et~al.(2018)Xu, Hu, Leskovec, and Jegelka]{xu2018powerful}
K.~Xu, W.~Hu, J.~Leskovec, and S.~Jegelka.
\newblock How powerful are graph neural networks?
\newblock \emph{arXiv preprint arXiv:1810.00826}, 2018.

\bibitem[Yang et~al.(2015)Yang, tau Yih, He, Gao, and Deng]{yang2015embedding}
B.~Yang, W.~tau Yih, X.~He, J.~Gao, and L.~Deng.
\newblock Embedding entities and relations for learning and inference in knowledge bases, 2015.

\bibitem[Zhang et~al.(2019)Zhang, Yao, Sun, and Tay]{zhang2019deep}
S.~Zhang, L.~Yao, A.~Sun, and Y.~Tay.
\newblock Deep learning based recommender system: A survey and new perspectives.
\newblock \emph{ACM computing surveys (CSUR)}, 52\penalty0 (1):\penalty0 1--38, 2019.

\bibitem[Zheng et~al.(2020)Zheng, Song, Ma, Tan, Ye, Dong, Xiong, Zhang, and Karypis]{DGL-KE}
D.~Zheng, X.~Song, C.~Ma, Z.~Tan, Z.~Ye, J.~Dong, H.~Xiong, Z.~Zhang, and G.~Karypis.
\newblock Dgl-ke: Training knowledge graph embeddings at scale.
\newblock In \emph{Proceedings of the 43rd International ACM SIGIR Conference on Research and Development in Information Retrieval}, SIGIR '20, page 739–748, New York, NY, USA, 2020. Association for Computing Machinery.

\bibitem[Zheng et~al.(2017)Zheng, Noroozi, and Yu]{zheng2017joint}
L.~Zheng, V.~Noroozi, and P.~S. Yu.
\newblock Joint deep modeling of users and items using reviews for recommendation.
\newblock In \emph{Proceedings of the tenth ACM international conference on web search and data mining}, pages 425--434, 2017.

\end{thebibliography}

\end{document}